\documentclass[twocolumn,superscriptaddress,english,prx,showpacs,longbibliography]{revtex4-2}

\usepackage{xcolor}
\usepackage{makecell}
\usepackage{enumitem} 
\usepackage{graphicx}
\usepackage{dcolumn}
\usepackage{bm}
\usepackage{amsmath,amssymb,amsthm,mathrsfs,amsfonts,dsfont}
\usepackage{balance}
\usepackage{array}
\usepackage{tabularx}
\usepackage{longtable}
\usepackage{indentfirst}
\usepackage{booktabs}
\usepackage{multirow}
\usepackage{placeins}
\usepackage{verbatim}
\usepackage{algorithm} 
\usepackage{algpseudocode}
\usepackage{physics}
\usepackage{braket}

\usepackage[columnwise]{lineno}

\usepackage[colorlinks=true,urlcolor=blue,citecolor=blue,linkcolor=blue]{hyperref}
\newcommand{\tP}{\mathcal{P}}
\newcommand{\tT}{\mathcal{T}}
\newcommand{\ri}{\partial i}

\begin{document}

\title{Tensor network dynamical message passing for epidemic models}

\author{Cheng Ye}
\affiliation{ CAS Key Laboratory for Theoretical Physics, Institute of Theoretical Physics, Chinese Academy of Sciences, Beijing 100190, China }
\affiliation{ School of Physical Sciences, University of Chinese Academy of Sciences,\- Beijing 100049, China }

\author{Zi-Song Shen}
\affiliation{ CAS Key Laboratory for Theoretical Physics, Institute of Theoretical Physics, Chinese Academy of Sciences, Beijing 100190, China }
\affiliation{ School of Physical Sciences, University of Chinese Academy of Sciences,\- Beijing 100049, China }

\author{Pan Zhang}
\email{panzhang@itp.ac.cn}
\affiliation{ CAS Key Laboratory for Theoretical Physics, Institute of Theoretical Physics, Chinese Academy of Sciences, Beijing 100190, China }
\affiliation{ School of Physical Sciences, University of Chinese Academy of Sciences,\- Beijing 100049, China }
\affiliation{School of Fundamental Physics and Mathematical Sciences,\- Hangzhou Institute for Advanced Study, UCAS, Hangzhou 310024, China}
\affiliation{Beijing Academy of Quantum Information Sciences, Beijing 100193, China}

\begin{abstract} 
While epidemiological modeling is pivotal for informing public health strategies, a fundamental trade-off limits its predictive fidelity: exact stochastic simulations are often computationally intractable for large-scale systems, whereas efficient analytical approximations typically fail to account for essential short-range correlations and network loops.
Here, we resolve this trade-off by introducing Tensor Network Dynamical Message Passing (TNDMP), a framework grounded in a rigorous property we term \textit{Susceptible-Induced Factorization}.
This theoretical insight reveals that a susceptible node acts as a dynamical decoupler, factorizing the global evolution operator into localized components.
Leveraging this, TNDMP provides a dual-mode algorithmic suite: an exact algorithm that computes local observables with minimal redundancy on tractable topologies and a scalable and tunable approximation for complex real-world networks.
We demonstrate that widely adopted heuristics, such as Dynamical Message Passing (DMP) and Pair Approximation (PA), are mathematically recoverable as low-order limits of our framework.
Numerical validation in synthetic and real-world networks confirms that TNDMP significantly outperforms existing methods to predict epidemic thresholds and steady states, offering a rigorous bridge between the efficiency of message passing and the accuracy of tensor network formalisms.
\end{abstract}

\maketitle

\section*{Introduction}
Infectious diseases have long posed a pervasive threat to humanity, from the catastrophic devastation of the Black Death~\cite{wade2020black,dean2018human} to the global disruption of the COVID-19 pandemic~\cite{bertozzi2020challenges}. To analyze and forecast these outbreaks, mathematical modeling has served as a cornerstone of epidemiology since Bernoulli’s study in 1760~\cite{bernoulli1760}, later maturing through the seminal Kermack-McKendrick~\cite{kermack1927contribution} and Reed-Frost~\cite{reed1928frost} models. The compartmental models, exemplified by the classic susceptible-infected-recovered (SIR) model~\cite{RevModPhys.87.925}, characterize epidemic processes by assuming that each individual in a population occupies one of several distinct states at any given time. These models have emerged as the prevailing paradigm due to their multifaceted analytical capacity, which enables the simulation of epidemic trajectories~\cite{MPBP,rDMP}, the reconstruction of infection origins~\cite{SIR-DMP,VAN_origin}, the derivation of critical epidemic thresholds~\cite{PhysRevLett.95.108701, PhysRevE.85.056111}, and the optimization of intervention strategies~\cite{Cohen2003,mazzoliSpatialImmunizationAbate2023}. Furthermore, owing to the universal nature of spreading dynamics, these frameworks have transcended their biological origins. They now underpin a vibrant interdisciplinary field, proving pivotal in areas as diverse as information dissemination~\cite{10.1038/s41598-020-62585-9,Rumour}, social behavior~\cite{centola2010spread,granovetter1978threshold}, financial contagion~\cite{haldane2011systemic}, and cascading failures in power grids~\cite{dobson2007complex,buldyrev2010catastrophic}.

A principal methodology in epidemic modeling involves constructing and solving dynamical equations. Analytical techniques, such as mean-field (MF) approaches~\cite{MeanField_early, HMF}, typically assume homogeneous mixing or predefined contact patterns, and  such approximations often fail in real-world settings where contact structure is crucial. This limitation has spurred the development of \textit{network epidemiology}~\cite{Networks,SIR-DMP,SIR-MP,rDMP,CME}, which models contacts explicitly as networks and shifts analysis from population densities to node-specific observables. However, exact equations in network epidemiology are hindered by long-range correlations, which generate intractable high-order terms. In practice, these terms are truncated via approximations that inevitably introduce error. Common methods include explicit techniques like Pair Approximation (PA)~\cite{Networks} or implicit ones like Dynamical Message Passing (DMP)~\cite{SIR-DMP,SIR-MP,rDMP}, which employs ``messages" that are also efficient nearest-neighbor approximations~\cite{CME,PhysRevE.89.022808}. Both constitute approximations within nearest-neighbor that ignore network loops, relying instead on tree-like assumptions and incurring a distinct loop-induced error. Efforts to correct these errors remain limited~\cite{LrDMP}. Although Monte Carlo simulations provide a robust benchmark, their computational cost, requiring extensive sampling for convergence in large systems, poses a significant bottleneck.

Tensor Network (TN) methods~\cite{ORUS2014117, SCHOLLWOCK201196} provide a powerful framework for encoding network structure and high-dimensional correlations, with proven success across various dynamical models~\cite{MPO_time_evolution, MPS_time_KCM, MPS_LD_KCM, PEPS_LD_KCM, PhysRevB.100.125121} and simple epidemic system~\cite{MPS_SIS}. However, TN simulations face a computational bottleneck at the global scale in epidemic models on complex network. For instance, a recent implementation using Matrix Product States~\cite{MPS_simulation_SIS} for epidemic modeling incurred costs exceeding those of classical heuristics for comparable accuracy. This inefficiency persists despite TN’s compression capabilities, highlighting a fundamental mismatch: explicitly tracking the full system state imposes unnecessary overhead for tasks like computing local marginals. It suggests that high precision does not require retaining full system-wide information, motivating the development of methods that avoid global evolution in favor of localized inference.

In this work, we introduce \textit{Tensor Network Dynamical Message Passing} (TNDMP), a method founded on a tensor network representation for epidemic dynamics, designed to overcome the aforementioned limitations. TNDMP fully accounts for correlations and network loops while using only the computational effort necessary to compute local observables. The approach is built on a rigorous theoretical foundation we term \textit{Susceptible-Induced Factorization}, which not only derives TNDMP but also unifies classical heuristics, such as Pair Approximation and standard DMP, as exact solutions on tree-like topologies. Furthermore, to address the scalability challenges posed by large, complex networks, we further introduce an approximation scheme controlled by a tunable parameter $N$, enabling TNDMP to handle short loops and short-range correlations rigorously via tensor network contraction, while treating long-range correlations through message passing.

We evaluate TNDMP on the SIR model across a range of networks, calculating single-node marginal probabilities. Its performance is compared against Pair Approximation (PA), standard Dynamical Message Passing (DMP), and a ground-truth benchmark from Monte Carlo (MC) simulation. To validate the method, we first apply exact TNDMP to synthetic networks containing artificial cliques—structures known to challenge classical approximations. On these benchmarks, TNDMP reproduces the MC results exactly. 
We further test the $N$-parameterized approximation of TNDMP on both synthetic and real-world contact networks. 
Even with a minimal refinement ($N=3$), TNDMP consistently outperforms PA and DMP, with accuracy improving as $N$ increases. In real-world networks with widespread epidemic, we observe a burn-out phenomenon, where predictions from classical methods exhibit significant dynamical deviations that exceed typical static errors. TNDMP effectively mitigates this issue, reducing both the overestimation of infection prevalence and the temporal shift in the predicted outbreak peak compared to existing heuristics. More broadly, the underlying tensor network framework and the TNDMP formulation are highly extensible, supporting future applications such as modeling temporal evolution and enabling reverse-time inference for a wide class of non-recurrent dynamical systems beyond epidemiology.

\section*{Results}

\subsection*{Epidemic Dynamics on Tensor Networks}
\begin{figure*}[!htb]
    \centering
    \includegraphics[width=\linewidth]{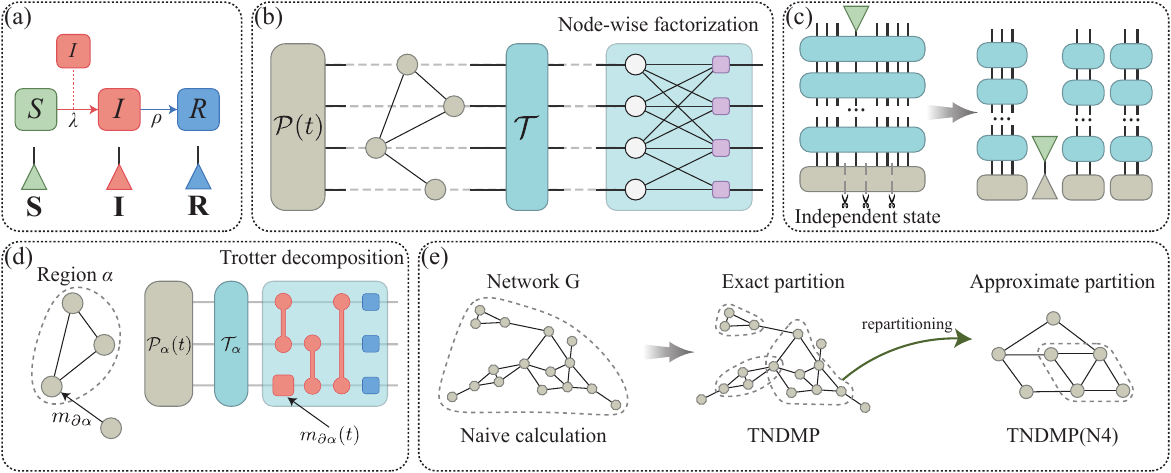}
    \caption{\textbf{Schematic illustration of tensor network dynamical message passing. }
(a) The Susceptible-Infected-Recovered (SIR) model, featuring three states and two transition processes. The three states are encoded as three vectors.
(b) The tensor $X_G^t$ with $n$ indices representing the probability of $3^n$ states in an $n$-node system. The state tensor evolves by contracting a tensor operator, which can be decomposed to local evolution of each node.
(c) Susceptible-Induced Factorization: contracting projector $S$ factorizes the global operator into sub operators with projector $S$ itself and consequently contract onto initial state at last. With an independent initial state, the joint probability given a node susceptible can be split into independent parts.
(d) Updating the tensor of a local area by contracting local operator with messages passed in, which only establishes on specific area based on the factorization property. 
(e) Illustration of exact and approximate partition, where the later is a repartition on each exact region. Obtained regions are circled by gray dashed lines, where the naive calculation is equivalent to assigning entire network as a region.}
    \label{fig:1}
\end{figure*}

The canonical SIR model~\cite{RevModPhys.87.925} serves as the simplest non-trivial framework for describing epidemic dynamics. Within this compartmental model, as illustrated in Fig.~\ref{fig:1}(a), the population is stratified into three distinct states: susceptible (S), infected (I), and recovered (R). The system evolves through two fundamental mechanisms: infection, in which a susceptible individual transitions to the infected state following exposure to an infected host; and recovery, characterized by the transition of an infected individual to the recovered state. These processes are formally represented as:
\begin{equation}\label{eq:evolve}
\begin{split}
SI & \xrightarrow{\lambda} II,\\
I & \xrightarrow{\rho} R,
\end{split}
\end{equation}
where $\lambda$ and $\rho$ denote constant rates of infection and recovery, respectively.

\subsubsection*{Tensor network form of epidemic dynamics}

For an SIR model defined on a graph $G = (V,E)$ with $n = |V|$ nodes, each node $i$ occupies one of the three discrete states $X_i \in \{S,I,R\}$, resulting in a total of $3^n$ possible system configurations $\mathbf{X} = (X_1,X_2,\dots,X_n)$. 
The distribution of joint probabilities $P(\mathbf{X},t)$ of global configurations at time $t$ is encoded as an $n$-index tensor: 
\begin{equation}
    \tP(t) = \sum_{\mathbf{X}} P(\mathbf{X},t) \ket{X_1} \ket{X_2} \dots \ket{X_n},
\end{equation}
where a vector $\ket{X_i}$ is one of the basis vectors depicted in Fig.~\ref{fig:1}(a). The temporal evolution of the state tensor is governed by the action of a transition operator:
\begin{align}
\tP(t+\tau) = \tT(\tau)\tP(t),
\end{align}
where $\tT(\tau)$ acts as a $2n$-index tensor operator, as illustrated in Fig.~\ref{fig:1}(b). For brevity, we eliminate the explicit time-step dependence and denote the operator as $\tT$.
Consequently, the state of the system at time $t$ is determined by the sequential application of the $t/\tau$ operator layers to the initial state $\tP(0)$:
\begin{equation}
    \tP(t) = \tT^{t/\tau} \tP(0). 
\end{equation}

A central theoretical finding within this framework is the ``Susceptible-Induced Factorization'' inherent to the SIR model's dynamics, which is derived from a node-wise factorization of $\tT$. It factorizes the $2n$-index tensor $\tT$ into a tensor network composed of single-node temporal operators and diagonal tensors as generalized Kronecker deltas. In the example shown in Fig.~\ref{fig:1}(b), the purple squares and white circles denote single-node operators and diagonal tensors, respectively, with the network topology encoded in the connections between these tensors.

Our analysis reveals that the susceptible projector, a vector of the susceptible state acting as a future state of a node $i$, denoted $\bra{S_i}$, possesses a unique algebraic property:
\begin{equation}\label{eq:factorize}
\bra{S_i} \tT = \prod_{\mathcal{C}\in \mathbf{C}(G\backslash i) } \tT_{\mathcal{C}}^i \bra{S_i}. 
\end{equation}
The contraction leaves $\bra{S_i}$ invariant and factorizes the operator $\tT$ into a product of independent components $\tT_{\mathcal{C}}^i$, each corresponding to a connected component $\mathcal{C}$ of the cavity graph $G \backslash i$ obtained by removing node $i$ from $G$. This mechanism effectively breaks the inter-neighbor correlations established by $\tT$; as depicted in Fig.~\ref{fig:1}(c), the contraction of $\bra{S}$ with the dynamical operator visually ``cuts'' the network connections at that node.

More broadly, this property propagates backward in time, implying that a susceptible projector serves as a ``breakpoint'' in the network's correlational structure throughout the epidemic history. Given an uncorrelated initial state, the joint probability distribution tensor with node $i$ susceptible $\bra{S_i}\tP(t)$ is factorized into several uncorrelated components, as shown in Fig.~\ref{fig:1}(c).

This factorization insight provides a theoretical foundation for developing efficient localized computational strategies. 
It reveals that in exact calculation of local observables, the dependence on the external environment inducing intractable global state evolution can be efficiently simplified and the factors corresponding to connected components $\mathcal{C}$ reveal the scope of application. 
On tree graphs, it leads to a derivation of pair approximation function and dynamical message passing equations presented in the Supplementary Materials.
Moreover, this critical insight directly underpins the development of our efficient algorithm for extracting local observables.

\subsubsection*{Tensor network dynamical message passing}

As illustrated in Fig.~\ref{fig:1}(d), the Susceptible-Induced Factorization facilitates the determination of local states. Specifically, a local state tensor $\tP_{\alpha}(t)$ is updated via the following contraction:
\begin{equation}\label{eq:local_update}
    \tP_{\alpha}(t+\tau) = \tT_{\alpha}(m_{\partial\alpha}(t)) \tP_{\alpha}(t),
\end{equation}
where the effective operator $\tT_{\alpha}$ integrates ``messages'', $m_{\partial\alpha}(t) = \{\sum_{j\in (\ri\cap\partial \alpha)}m_{j\to i}(t)| i \in \alpha\}$, originating from the external environment. A message $m_{j \to i}(t)$ is the conditional probabilities $P(I_j|S_i,t)$ that node $j$ is infected given that node $i$ is susceptible at time $t$.
This local update scheme is applied to single nodes and designated spatial domains, where ``regions'' are defined as the atomic elements, while other eligible areas are their combination. 

These regions, derived through network partitioning, collectively constitute a complete cover of the network. Consequently, any local observables that depend directly on the state of an area $\tP_A(t)$ can be precisely evaluated using the region state $\tP_\alpha(t)$ where $A\subseteq\alpha$ (or a combination thereof) through local updates, thus obviating the computational necessity of resolving the global state $\tP(t)$ at every timestep.
In particular, these local updates are based on message exchanges across edges associated with adjacent regions. Therefore, the state of the region encompassing the edge $(i,j)$ is necessary to evaluate $m_{j \to i}(t)$, which subsequently requires further messages and region states. Ultimately, this framework accounts for the states across all regions, but maintains superior computational efficiency compared to the direct simulation of the global state.

Leveraging the concept of local updates, we developed the Tensor Network Dynamical Message Passing (TNDMP) algorithm to efficiently evaluate local observables.
The TNDMP algorithm initially identifies the reducible structure of the system by partitioning the network into regions, as illustrated in Fig.~\ref{fig:1}(e). 
Subsequently, the states of all regions and combinations required for multi-region-spanning observables are computed step by step via tensor contraction according to Eq.~\eqref{eq:local_update}. 
In doing so, TNDMP simultaneously derives the messages $m_{j \to i}(t) = P(I_jS_i,t)/P(S_i,t)$ required by operators $\tT_{\alpha}(m_{\partial\alpha}(t))$. Here, $P(I_jS_i,t)$ is obtained by contracting projectors on $\tP_{\alpha}$ during each iteration, while $P(S_i,t)$ and the single-node tensors$\tP_{i}(t)$ are calculated separately through local update contractions with $\tT_{i}(m_{\ri}(t))$.
Our method maintains mathematical rigor by introducing no approximations beyond those implicit in the network partitioning itself. Consequently, TNDMP yields a lower computational overhead compared to the direct and exact simulation of the global state.

To further reduce computational complexity during contraction, we introduce an additional decomposition of the operator $\tT_{\alpha}$, which is distinct from the node-wise decomposition shown in Fig.~\ref{fig:1}(b). The implemented Trotter decomposition, illustrated in Fig.~\ref{fig:1}(d), utilizes isolated single-node operators and pairwise operators, which are analogous to single bit gates and two bit gates, to model infection and recovery processes. Specifically, infections originating from external nodes are modeled by single-node gates conditioned on incoming messages, while internal infections are represented by pairwise operators.

Moreover, because the state tensor of an $n$-node region scales as $3^n$, this exponential growth renders exact calculations for large regions computationally prohibitive. To address this scalability challenge, we propose an approximation scheme that subdivides oversized exact regions into approximate subregions of manageable size. We define a parameter $N$ to represent the maximum allowable size for these approximate regions. This approach can be conceptualized as a hierarchical sub-partitioning applied to each exact region (see Fig.~\ref{fig:1}(e)). The approximation method is denoted by its parameter value, for instance, ``$N4$'' in Fig.~\ref{fig:1}(e) signifies a partition where $N=4$. In the limit where $N=2$, regions are restricted to individual edges, causing the TNDMP algorithm to effectively reduce to a pair approximation formulated within a tensor network framework.

\subsubsection*{Algorithm}

The pseudocode for calculating the marginals $\tP_i$ by the TNDMP is summarized below:
\begingroup
\setlist{nosep}
\begin{itemize}
    \item \textbf{Input}: Network $G=(V,E)$, partition parameter $N$, initial information (patient zero or infected probabilities), epidemic parameters $\lambda, \rho$, time step $\tau$, end time $T$.
\end{itemize}
\begin{enumerate}
    \item \textbf{Network Partitioning}: Divide network $G$ into a set of regions $\mathcal{R}$ by partitioning with the parameter $N$.
    \item \textbf{Initialization}: Initialize messages $\{m_{i \to j}(0),(i,j) \in E\}$, state tensor of regions $\{\tP_\alpha(0), \alpha \in \mathcal{R}\}$ and nodes $\{\tP_{i}(0),i \in V\}$ by initial information.
    \item \textbf{Iterative Evolution}: For each time step $t = 0, \tau, \dots, T-\tau$:
    \begin{enumerate}
        \item \textbf{Region Update}: Update each state tensor $\tP_\alpha(t)$ by contracting Trotter operators that are factorized from operator $\tT_{\alpha}(m_{\partial\alpha}(t))$;
        \item \textbf{Node Update}: Update each $\tP_{i}(t)$ by contracting the operator $\tT_{i}(m_{\ri}(t))$;
        \item \textbf{Message Passing}: Apply projectors to $\tP_\alpha(t+\tau)$ and $\tP_{i}(t+\tau)$ to extract the joint and marginal probabilities, $P(I_jS_i,t+\tau)$ and $P(S_i,t+\tau)$, respectively. Calculate the new messages as: $m_{i \to j}(t+\tau) = P(\mathrm{I}_j \mathrm{S}_i, t+\tau)/P(\mathrm{S}_i, t+\tau)$,$ \forall (i,j) \in E.$
    \end{enumerate}
\end{enumerate}
\begin{itemize}
    \item \textbf{Output}: Marginals $\{\tP_{i}(t)| i \in V \}$ for $t = 0,\tau,2\tau,\dots,T$.
\end{itemize}
\endgroup

\begin{figure*}[!htb]
    \centering
    \includegraphics[width=\linewidth]{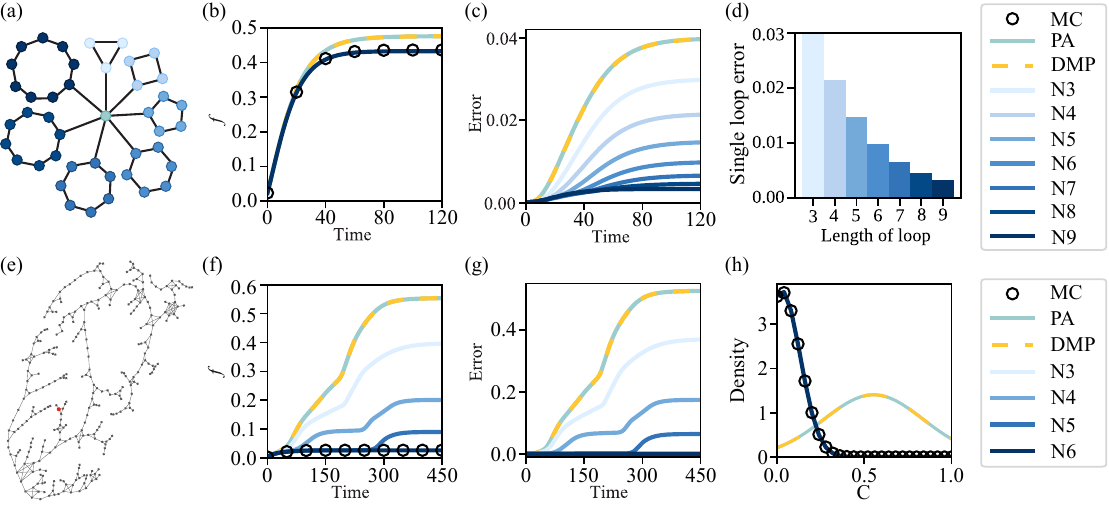}
    \caption{\textbf{Performance evaluation on synthetic networks with feasible regions.} 
    (a) A network consisting of $n = 43$ nodes with an ideal exact region partition. The structure contains one intermediate node and seven loops with sizes ranging from 3 to 9. The intermediate node is patient zero, representing the only infectious individual at $t=0$. 
    (b) Temporal evolution of the cumulative infection fraction $f = \frac{1}{n} \sum^{n-1}_{i=0} P(C_i)$, where $P(C_i) = P(I_i) + P(R_i)$. We compare the ground-truth marginals derived via Monte Carlo (MC) simulations with Pair Approximation (PA), Dynamical Message Passing (DMP), and our rigorous method ($N9$). 
    (c) The error in cumulative infection marginals, $E = \frac{1}{n}\sum_{i=0}^{n-1} |P(C_{i}) - P^{MC}(C_i)|$, obtained by our approximation for all non-trivial values of $N \in [3, 9]$, as well as PA and DMP. 
    (d) Final errors contributed by single loops, arranged according to their lengths. 
    (e) A network of $n=313$ nodes generated from a random tree augmented with cliques of sizes 3 to 6; the respective counts for these cliques are [35, 10, 2, 1]. The experiment originates from a single patient zero (colored red). 
    (f) The fraction of cumulative infection $f$ on network in panel (e) over time. 
    (g) Error in cumulative infection marginals through time for existing algorithms compared to our approximation with various $N$. 
    (h) Probability density distribution of cumulative infection marginals at the steady state, comparing rigorous TNDMP ($N6$) with other algorithms.}
    \label{fig:2}
\end{figure*}

\subsection*{Applications of TNDMP}
In our experiments, the SIR model is parameterized with an infection rate of $\lambda = 0.1$ and a recovery rate of $\rho = 0.05$. The results obtained from $10^6$ Monte Carlo (MC) simulations serve as the ground truth. For the majority of our evaluations, we report the marginals of the cumulative infection $C$, defined as the sum of the probabilities of being in the infected ($I$) and recovered ($R$) states.

\subsubsection*{Experiment on synthetic networks}

We initiate our evaluation using networks with tractable exact partitions, where the exact regions requiring local tensor network contractions remain computationally manageable. This setup enables TNDMP to achieve high accuracy by employing a sufficiently large, yet computationally feasible, parameter $N$. 
Initially, TNDMP is validated on a network topology containing multiple disjoint loops with lengths ranging from 3 to 9, as illustrated in Fig.~\ref{fig:2}(a). As $N$ increases, our approximation systematically incorporates these loops into the partitioning in ascending order of their lengths. By designating the central node as the only initially infected node (patient zero), the epidemic dynamics are restricted to individual loops. This configuration ensures that error propagation between disjoint loops is prevented.
Consequently, the error variance across different $N$ values directly reflects the specific error contributions of single loops relative to their lengths. Fig.~\ref{fig:2}(b) displays the cumulative infection fraction $f = \frac{1}{n}\sum^{n-1}_{i=0} P(C_{i})$, where $P(C_i) = P(I_i)+P(R_i)$ represents the marginal cumulative infection probability for node $i$. We compare the ground truth from MC simulations with results from PA and DMP, alongside exact TNDMP approach ($N=9$ in this network). Our method demonstrates high fidelity by closely tracking the MC results, whereas PA and DMP exhibit significant deviations.

Fig.~\ref{fig:2}(c) illustrates the $L_1$ error for TNDMP across $N$ ranging from 3 to 9, compared to PA and DMP. The results indicate that the TNDMP error decreases monotonically as $N$ increases, where PA can be considered equivalent to TNDMP with $N=2$ as previously noted. The residual error at $N=9$ is minimal and primarily attributable to the Trotter decomposition. Furthermore, this experimental design allows us to quantify the error contributions inherent to single loops of varying lengths (neglecting the Trotter-induced bias). For clearer illustration, these loop-specific errors are presented at the steady state ($t=150$) in Fig.~\ref{fig:2}(d). The results reveal that the error contribution from a single loop is inversely correlated with its length. This observation supports the hypothesis that longer loops induce smaller errors, which serves as a fundamental motivation for our approximation method, even though the current experiment focuses on independent loops.

\begin{figure*}[!htb]
    \centering
    \includegraphics[width=0.92\linewidth]{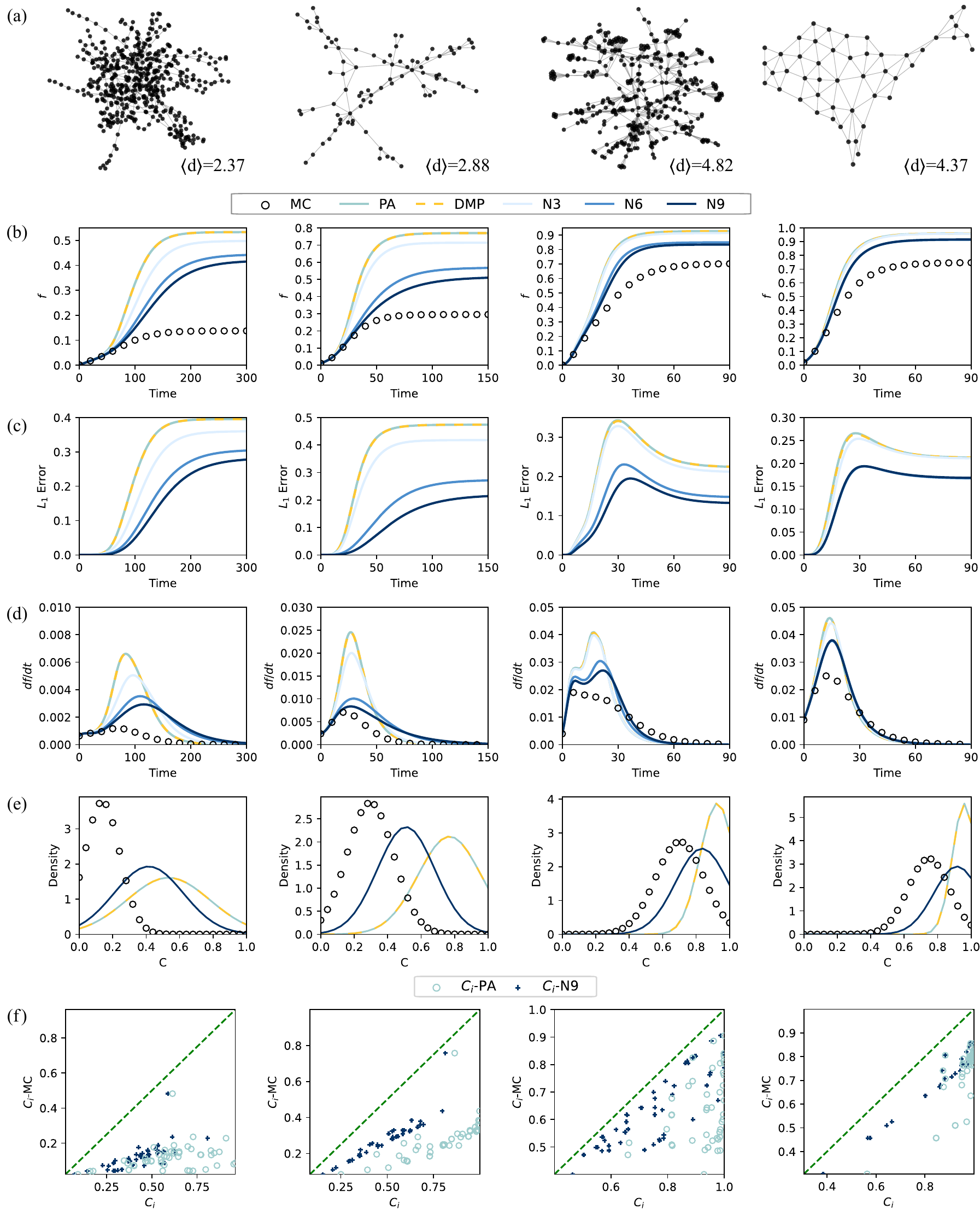}
    \caption{\textbf{Comparative analysis of epidemic dynamics and steady-state distributions in real-world networks.} 
    Experimental results for four real-world networks are presented in columns, respectively. 
    (a) Structural sketches of the networks annotated with their average degrees $\langle k \rangle$. 
    (b) Temporal evolution of the cumulative infection fraction $f$. 
    (c) Temporal evolution of the $L_1$ error $E$ in cumulative infection marginals. 
    (d) Temporal evolution of the the derivative $df/dt$. 
    (e) Probability density function of the cumulative infection marginals at the steady state. 
    (f) Scatter plot of the steady-state marginal probabilities $C_i$ for 50 randomly selected nodes. The $y$-axis represents the MC ground truth, while the $x$-axis represents results obtained via TNDMP ($N=9$) and PA. The green dashed line indicates the ideal $y=x$ agreement with the ground truth.}
    \label{fig:3}
\end{figure*}

To evaluate the performance of our approximation, we constructed a globally tree-like network incorporating cliques. This was achieved by first generating a random tree and subsequently replacing each node of degree $d \ge 3$ with a $d$-node clique, where the original edges were remapped to the new clique vertices. While this construction poses a significant challenge for classical methods predicated on tree-like assumptions, its exact partition is unambiguous: each clique is defined as a single region, while the remaining edges in the tree are treated as individual regions. Consequently, the maximum cardinality of the cliques directly determines the value of the threshold parameter $N$ at which our methodology achieves exact results.
In our evaluation, we generated a 200-node random tree and replaced 48 nodes with cliques, the largest of which had a size of 6. This setup ensures that TNDMP achieves exact calculation at $N = 6$. A schematic of this network, with patient zero colored red, is illustrated in Fig.~\ref{fig:2}(e). Panels (f) and (g) of Fig.~\ref{fig:2} present the cumulative infection fraction $f$ and its $L_1$ error on this network, comparing our approximation for $N \in \{3,4,5,6\}$ against PA and DMP. For $N < 6$, portions of the cliques are pruned to satisfy the $N$ constraint. The results exhibit a monotonic decrease in TNDMP error as $N$ increases, with the rigorous $N=6$ algorithm yielding a Trotter error below $10^{-3}$. Notably, all our approximations, even at $N=3$, yield significantly lower errors than PA and DMP.

Furthermore, Fig.~\ref{fig:2}(h) illustrates the probability density function of the cumulative infection at the final time step. The steady states reveal a substantial disparity between the ground truth obtained from Monte Carlo (MC) simulations and the endemic states predicted by tree-based methods. In contrast, the results from our method closely align with the MC simulations.

\subsubsection*{Applications on real-world networks  }
We further evaluated the approximate TNDMP method on four real-world networks for which a rigorous TNDMP approach is computationally prohibitive: an electric power grid ($n=494$), a collaboration network of Sandia National Labs scientists ($n=86$), a co-authorship network ($n=379$), and the network of contiguous USA states ($n=49$)~\cite{nr}. Structural sketches of these networks are presented in Fig.~\ref{fig:3}(a).

The results presented in Fig.~\ref{fig:3} consistently demonstrate that our TNDMP method outperforms PA and DMP. Rows (b) and (c) in Fig.~\ref{fig:3} depict the temporal evolution of the cumulative infection fraction and the corresponding $L_1$ error $E$ for PA, DMP, and our approximations with $N \in \{3,6,9\}$. An inverse relationship between the error and the parameter $N$ is observed throughout the evolution.
A distinctive feature in the error profiles of the latter two networks is the occurrence of pronounced peaks followed by a transient reduction, despite the absence of substantial divergence in the cumulative infection fraction. To investigate this phenomenon, Fig.~\ref{fig:3}(d) presents the derivative of the cumulative infection fraction $df/dt$. As observed in the panels for the latter two networks, the derivative decreases rapidly following its peak, eventually falling below the levels recorded in Monte Carlo simulations. In contrast, for the first two networks, the derivatives predicted by the approximate methods remain consistently higher than the MC results.
In conjunction with the high cumulative infection fraction $f$, we attribute the subsequent reduction of errors to the following mechanism: in the latter two networks, the overestimation of the derivative by approximate methods is so pronounced that it leads to a rapid depletion of the susceptible population, triggering a premature \textbf{burn-out} of the epidemic. Thus, instead of sustaining a more enduring propagation, characterized by the broader $df/dt$ peaks seen in the first two networks, the majority of individuals are infected within a very short duration in the last two networks. Consequently, this burn-out effect causes $df/dt = \cfrac{\lambda}{n} \sum_{(i,j)\in E} P(S_iI_j)$ in the approximate simulations to weaken rapidly because few individuals remain susceptible, while the derivative in the ground-truth system, $df_{MC}/dt$, decreases more gradually. Since errors primarily stem from overestimated infection, such that $df/dt - df_{MC}/dt \approx dE/dt$, the crossover point where $df/dt$ falls below $df_{MC}/dt$ corresponds to the observed error peak followed by a transient decrease.

Fig.~\ref{fig:3} (e) and (f) focus on the steady-state results for TNDMP with $N=9$. The fifth column displays the probability density function of the final cumulative infection, while the last column compares the steady-state marginal probabilities of 50 randomly selected nodes between TNDMP ($N=9$) and PA. These density and marginal plots further corroborate our explanation of the error dynamics: the approximate simulations for the latter two networks converge to a state of total \textbf{burn-out}, where most individuals reach a cumulative infection marginal near 1.0, whereas the true infected fraction is significantly smaller. Although TNDMP also exhibits a tendency to overestimate infection, it consistently provides results superior to PA and DMP. The density plots reveal that our method yields a distribution closer to the ground truth, and the scatter plots confirm that the TNDMP results generally align more closely with the true values.
Across the four real-world networks analyzed, our method demonstrates its most significant performance gains on the second network. This observation aligns with our theoretical expectations, as TNDMP is specifically designed to yield superior results on topologies that are predominantly sparse and tree-like. In contrast, the advantage over PA and DMP is less pronounced in the final lattice-like network, which presents a significant challenge for all methods due to its high clustering.
In the third network, we observe two distinct peaks in the $df/dt$ profile. This phenomenon mirrors the stepped behavior previously noted in Fig.~\ref{fig:2}(f), which corresponds to the impact of numerous short loops that are neglected when sub-partitioning densely connected cliques. These peaks highlight the substantial influence of clustered short loops on epidemic dynamics and underscore the potential for further performance gains if clique structures are managed more precisely within our framework.

\subsubsection*{Computational Efficiency and Scalability}

To demonstrate the necessity of the proposed $N$-based approximation and evaluate its computational cost, we analyze the 494-bus network. This topology presents a significant challenge for exact inference owing to its loop-rich structure. Specifically, the largest region requiring rigorous treatment comprises 212 nodes. In the context of the $SIR$ model, computing the exact marginals for this region would require summing over a state space of dimension $3^{212} \approx 1.7 \times 10^{101}$, rendering the exact solution computationally intractable. However, the $N=9$ approximation partitions the system into 33 multi-edge regions within a negligible duration (less than 1 second) and reduces the state space from $3^{212}$ to less than $33 \times 3^9$ region-states plus the minor state spaces of remaining individual edges.

Consequently, efficient approximation schemes are essential. Table~\ref{tab:runtime} compares the CPU running time of the proposed TNDMP method against the Monte Carlo (MC) baseline and tree-like approximations (PA and DMP). While MC simulations ($10^6$ realizations) provide a statistical baseline, they incur a computational cost three orders of magnitude higher than deterministic methods.

\begin{table}[h]
    \centering
    \caption{\textbf{Comparison of CPU Running Time on the 494-Bus Network.} The exact solution is infeasible ($3^{212}$ states). TNDMP ($N=9$) achieves a runtime comparable to fast heuristics (PA, DMP) while avoiding the heavy computational burden of Monte Carlo (MC) simulations. The reported time for TNDMP ($N=9$) refers to the dynamic simulation duration. The preprocessing time for network partitioning is less than 1 second and is not included in the simulation runtime.}
    \label{tab:runtime}
    \resizebox{\linewidth}{!}{
    \begin{tabular}{lcc}
        \toprule
        \textbf{Method} & \textbf{Time (sec.)} & \makecell[c]{\textbf{Relative}\\ \textbf{Cost}} \\
        \midrule
        PA & 13.06 & $1.0\times$ \\
        DMP & 13.10 & $\approx 1.0\times$ \\
        \textbf{TNDMP ($N=9$)} & 212.18 & $\mathbf{\approx 16.3\times}$ \\
        MC & 137073.54 & $> 10,000\times$ \\
        Exact & -- & \textit{Intractable} ($3^{212}$) \\
        \bottomrule
    \end{tabular}
    }
\end{table}

Significantly, TNDMP ($N=9$) achieves a critical balance between rigor and efficiency. Despite incorporating higher-order correlations to address the inaccuracies, its computational cost exceeds that of the ultra-fast DMP and PA heuristics by a factor of only approximately 16. This result indicates that TNDMP successfully bridges the gap between accuracy and efficiency, offering near-exact precision at a cost comparable to classical heuristics.

\section*{Discussion}
We introduce a tensor network (TN) framework for characterizing the dynamics of epidemic models on networks. This formalism provides a robust mathematical toolkit for analytical derivation and naturally unifies classical Dynamical Message Passing (DMP) and Pair Approximation (PA) as exact methods for tree topologies. The framework encompasses Tensor Network Dynamical Message Passing (TNDMP) with both an exact methodology and efficient approximation techniques. Our approach facilitates the determination of local observables with significantly enhanced accuracy at controllable computational costs.

The primary source of error in our algorithm stems from approximating loop structures to reduce computational complexity. This approximation is necessitated by the exponential growth of state tensors. While representing the global state remains challenging, tensor network methods such as Matrix Product States (MPS)~\cite{ORUS2014117}, Projected Entangled Pair States (PEPS)~\cite{PEPS}, and Tree Tensor Networks (TTN)~\cite{TTN} hold promise for the efficient representation of local region-state tensors. Instead of sub-partitioning oversized regions, representing large region tensors via TN ansatzes offers a pathway to higher accuracy, and a hybrid approach combining both techniques is also feasible.

We currently employ a greedy heuristic for the $N$-based approximation. Consequently, the resulting partition is not guaranteed to be optimal and may not minimize the approximation error. This limitation can be addressed by formulating the partitioning as a formal optimization problem and employing advanced combinatorial optimization algorithms. Additionally, in the Supplementary Materials, we introduce an alternative heuristic scheme, which serves as a supplement to the primary $N$-based approximation.

Although the current implementation focuses on algorithmic accuracy on CPUs, the TNDMP framework naturally lends itself to parallel computing and hardware acceleration. The local update operations (Eq.~\ref{eq:local_update}) for disjoint regions are independent and can be executed simultaneously. Furthermore, tensor contraction, which governs the computational cost, can be significantly accelerated using modern Graphics Processing Units (GPUs) or Tensor Processing Units (TPUs). Future implementations leveraging these hardware capabilities could reduce runtime, enabling the rigorous analysis of significantly larger social and technological networks.

Finally, the present approach exhibits high extensibility to variants of the SIR model, such as the SEIR model~\cite{hethcote2000mathematics} and rumor spreading~\cite{Rumour}. However, significant challenges remain for recurrent models, such as Susceptible-Infected-Susceptible (SIS) dynamics. In the absence of the Susceptible-Induced Factorization property, TNDMP reduces to a probabilistic approximation that lacks the rigorous physical decoupling inherent to the SIR framework. However, the tensor network formalism introduced in this paper provides a promising analytical foundation for characterizing complex dynamics and a powerful tool for representing correlations.

\section*{Methods}

\subsection*{Node-wise factorization over the temporal operator}

\begin{figure}[!htb]
\centering
\includegraphics[width=\linewidth]{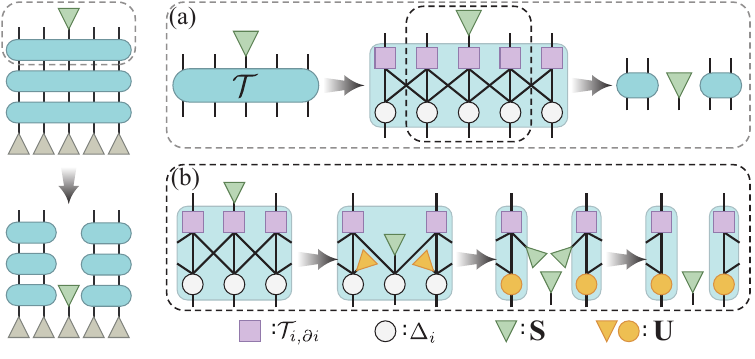}\hspace{3pt}
\caption{\textbf{The pictorial illustration of Susceptible-Induced Factorization on a 1-dimensional chain.} Sketch of the tensor contraction of a susceptible projector $\bra{S}$ and a state tensor $\tP(t)$. (a) The contraction breaks factorized $\tT$ into $\tT_{\mathcal{C}}^{i}$. (b) Pictorial function of contracting $\bra{S}$ onto factorized $\tT$.}
\label{fig:TN}
\end{figure}

We use a node-wise factorization to factorize $\tT$. It originates from a factorization of $T(\mathbf{X}^{t+\tau} | \mathbf{X}^t)$, the probability of the transition from a configuration $\mathbf{X}^t$ to another $\mathbf{X}^{t+\tau}$ over a time step $\tau$. By the locality of interactions in the epidemic models, this global probability is factorized over the neighborhoods of individual nodes:
\begin{equation}
    T(\mathbf{X}^{t+\tau} | \mathbf{X}^t) = \prod_{i \in V} T\left(X_{i}^{t+\tau} | X_i^t,X_{\ri}^t\right),
\end{equation}
where $\ri$ denotes the set of immediate neighbors of node $i$. We reformulate this factorization within a tensor network formalism. The global operator $\tT$ decomposes into a contraction of local tensors:
\begin{equation}
    \label{eq:operator}
    \tT = \prod_{i\in V} \tT_{i,\ri} \Delta_{i} ,
\end{equation}
Here, the single-node transition tensor $\tT_{i,\ri}$ encodes the local conditional probability $W(X_{i}^{t+\tau} | X_i^t,X_{\ri}^t)$, acting on the state indices of node $i$ and its neighbors. The tensor $\Delta_{i}$, a generalized Kronecker delta, ensures the physical consistency of the state $X_i^t$ on the indices associated with the previous state of $i$. We show this factorization on a 1-dimensional chain in Fig.~\ref{fig:TN}(a)

\subsection*{Susceptible-Induced Factorization}

The computational tractability of our framework rests on a rigorous algebraic property of the SIR dynamics, the ``Susceptible-Induced Factorization''. Physically, if a node $i$ is known to be in the susceptible state at time $t$, its history effectively decouples the dynamical correlations between its neighbors.

Mathematically, this is realized by contracting the ``susceptible projector'' $\bra{S_i}$ onto the future index of the global evolution operator $\tT$. This contraction propagates through the network structure, factorizing $\tT$ into disconnected components. Specifically, contracting $\bra{S_i}$ with the local transition tensor $\tT_{i,\ri}$ yields:
\begin{equation}\label{eq:SP}
    \bra{S_i}\tT_{i,\ri} =  \prod_{j \in \ri} \bra{U_j} \bra{S_i},
\end{equation}
where we have employed the first-order approximation in $\tau$, which can be sufficiently small in the derivation. The vector $\bra{U} = \bra{S} + (1-\lambda\tau)\bra{I} +\bra{R}$ represents the constraint imposed on the neighbor $j$ by the fact that node $i$ remained susceptible.

Crucially, the diagonal tensor $\Delta_{i}$ copies the projector as a product of $\bra{S_i}$ to other indices. Consequently, the contraction $\bra{S_i} \tT$ shown in Fig.~\ref{fig:TN} efficiently removes node $i$ from the tensor network of factorized $\tT$, breaking the global operator into a tensor product of operators acting on the disconnected components of the cavity graph $G \backslash i$:
\begin{equation}
    \bra{S_i} \tT = \prod_{\mathcal{C}\in \mathbf{C}(G\backslash i)} \tT_{\mathcal{C}}^{i} \bra{S_i} .
\end{equation}
Here, $\mathbf{C}(G\backslash i)$ denotes the set of connected components in the graph when node $i$ is removed. 
This factorization implies that for an uncorrelated initial state $\tP(0)$, the marginal probability of observing node $i$ as susceptible at time $T$, along with the system's conditional state, factorizes completely:
\begin{equation}
    \label{eq:PS}
        \bra{S_i}\tP(t) = P(S_i,0) \prod_{\mathcal{C} \in \mathbf{C}(G\backslash i)} \tP_{\mathcal{C}}^{i}(t).
\end{equation} 

With the tree-assumption, this ``Susceptible-Induced Factorization'' can immediately derive the key equation of Dynamical Message Passing by summarizing indices of $\bra{S_i}\tP(t)$ to obtain susceptible marginals:
\begin{equation}
    P(S_i,t) = P(S_i,0) \prod_{j \in \ri} H_{j}^{i}(t),
\end{equation}
where the scalar $H_{j}^{i}(t) = \sum_{\mathbf{X}_{\mathcal{C}_j}}\tP_{\mathcal{C}_j}^{i}(t)$ is the summation over all configurations of the subnetwork $\mathcal{C}_j$ is the DMP message for the SIR model. The labels of neighbors take the place of connected components, since in a tree network each neighbor $j$ corresponds to a unique connected component $\mathcal{C}_j$ of the cavity graph $G\backslash i$. Refer to Supplementary Materials for the derivation of the dynamical equation of $H_{j}^{i}(t)$.
Meanwhile, by keeping the state of two neighbors $j$ and $k$ of node $i$ fixed in $\bra{S_i}\tP(t)$, we obtain the pair approximation function:
\begin{equation}
P(X_jS_iX_k,t) = \cfrac{P(X_jS_i,t)P(S_iX_k,t)}{P(S_i,t)},
\end{equation}
where the intermediate node is constrained to be susceptible.

Most importantly, it allows us to replace the intractable cascading expansion with local message passing updates rigorously, as detailed in the Supplementary Materials. Specifically, as we have mentioned, for a region $\alpha$ where the boundary nodes satisfy specific connectivity constraints (see Eq.~\ref{eq:local_update}), the influence of the external network is fully captured by scalar messages $m_{\partial\alpha}(t)$, allowing the exact local update $\tP_{\alpha}(t+\tau) = \tT_{\alpha}(m_{\partial\alpha}(t)) \tP_{\alpha}(t)$.

\subsection*{Algorithm: Scalable $N$-Partitioning}

To reconcile the competing demands of accuracy and computational tractability, we introduce a hierarchical partitioning scheme governed by a tunable parameter $N$, representing the maximum allowable region size. This framework leverages the insight that while short-range loops necessitate exact tensor contraction, long-range dependencies can be effectively approximated via dynamical message passing. It is grounded in the decay of correlations with increasing loop length, a premise empirically validated by our experiments on synthetic topologies (Fig.~\ref{fig:2}(a-d)).

The algorithm proceeds recursively, initially decomposing the network into its fundamental biconnected components. For any component exceeding the size threshold $N$, we apply a topological ``peeling'' protocol that iteratively excises edges associated with the longest cycles until the component's core size satisfies the threshold condition. Crucially, the set of excised edges is not merely treated as residual. Any non-trivial biconnected subgraphs formed by these edges are recursively subjected to the same peeling protocol. Ultimately, the resulting dense cores are retained as tensor network regions for exact evolution, while the remaining edges serve as message-passing channels. This hybrid architecture ensures that the most significant correlations (short loops) are treated rigorously, while constraining the computational complexity of any region to $\mathcal{O}(3^N)$.

\subsection*{Time Evolution via Trotter Decomposition}

The direct contraction of the region operator $\tT_{\alpha}$ becomes computationally intensive as the region size grows. To optimize the local update step, we employ a first-order Trotter-Suzuki decomposition, decoupling the evolution into a sequence of sparse, local gate operations:
\begin{equation}
    \label{eq:Trotter}
    \tT_{\alpha}(m_{\partial\alpha}(t)) \approx \prod_{i\in \alpha} \mathbf{R}_{i} \prod_{(i,j)\in E_{\alpha}}\mathbf{I}_{ij} \prod_{i\in \alpha} \mathbf{I}_{i}(m_{\partial\alpha\to i}(t)).
\end{equation}
Here, $E_{\alpha}$ denotes the set of edges strictly contained within region $\alpha$. The term $m_{\partial\alpha\to i} = \sum_{j \in (\ri \cap \partial \alpha)} m_{j \to i}$ represents the aggregate infection pressure exerted on node $i$ by its neighbors in the external boundary $\partial \alpha$.

The operators in Eq.~\eqref{eq:Trotter} correspond to specific physical processes: $\mathbf{R}_{i}$ acts as a single-site recovery gate; $\mathbf{I}_{ij}$ represents the pairwise infection gate for internal edges; and $\mathbf{I}_{i}(m_{\partial\alpha\to i}(t))$ is a single-site gate conditioned on the external messages, effectively modeling the "field" produced by the environment. This decomposition reduces the update cost to a series of low-cost tensor contractions, enabling efficient simulation even for regions with dense internal connectivity.

\begin{acknowledgments}
A Python implementation of our algorithm and a Jupyter Notebook for reproducing the data of the figures are available at~\cite{code}.
We are grateful to Yijia Wang and Jing Liu for helpful discussions. We thank Dong-Yang Feng for his assistance with figure preparation.
The work is supported by Projects 12325501 and 12447101 of the National Natural Science Foundation of China.
\end{acknowledgments}

\bibliography{main}

\setcounter{equation}{0}
\setcounter{figure}{0}
\setcounter{table}{0}
\renewcommand{\theequation}{S\arabic{equation}}
\renewcommand{\thefigure}{S\arabic{figure}}
\renewcommand{\thetable}{S\arabic{table}}

\onecolumngrid

\begin{center}
\vspace{2em}
\textbf{SUPPLEMENTARY MATERIALS OF ``Tensor network dynamical message passing for epidemic models"}
\vspace{1em}
\end{center}

\section{Pair approximation and Dynamical message passing}
The standard dynamical equations of marginals of the SIR model on networks are:
\begin{align}
    \label{eq:ODE_PA}
    \cfrac{dP(S_i)}{dt} &= -\lambda \sum_{j\in \ri} P(S_{i}I_{j}), \\
    \cfrac{dP(I_i)}{dt} &= \lambda \sum_{j\in \ri} P(S_{i}I_{j}) - \rho P(I_i),\\
    \cfrac{dP(R_i)}{dt} &= \rho P(I_i),\\
    \cfrac{dP(S_iI_j)}{dt} &= -(\rho+\lambda) P(S_iI_j) + \lambda \sum_{k\in \partial_j \backslash i} P(S_iS_jI_k) - \lambda \sum_{k\in \partial_i \backslash j} P(S_iI_jI_k),
\end{align}
where the hinder of correlation appears as the triple terms which also require dynamical equations to evolve and expand to fourth-order terms. The function of pair approximation as the second moment closure method is:
\begin{equation}
    P(X_iX_jX_k) = \cfrac{P(X_iX_j)*P(X_jX_k)}{P(X_j)},
\end{equation}
where $i,k$ are neighbors of $j$. Applying on the last equation, we have:
\begin{equation}
    \cfrac{dP(S_iI_j)}{dt} = -(\rho+\lambda) P(S_iI_j) + \lambda P(S_iS_j)\sum_{k\in \partial_j \backslash i} \cfrac{P(S_jI_k)}{P(S_j)} - \lambda P(S_iI_j)\sum_{k\in \partial_i \backslash j} \cfrac{P(S_iI_k)}{P(S_i)}.
\end{equation}
Induces further $P(S_iI_j)$, whose dynamical equation with pair approximation is:
\begin{equation}
    \cfrac{dP(S_iS_j)}{dt} = - \lambda P(S_iS_j)\sum_{k\in \partial_j \backslash i} \cfrac{P(S_jI_k)}{P(S_j)} - \lambda P(S_iS_j)\sum_{k\in \partial_i \backslash j} \cfrac{P(S_iI_k)}{P(S_i)}.
\end{equation}
From the function, we can see that the pair approximation function only applies on terms with susceptible intermediate nodes. Thus, the actual function of PA applied in the SIR models is 
\begin{equation}
    P(X_iS_jX_k) = \cfrac{P(X_iS_j)*P(S_jX_k)}{P(S_j)}.
\end{equation}

The key of the Dynamical Message Passing (DMP) algorithm is the alternative equation of susceptible marginals:
\begin{equation}
    P(S_i,t) = z*\prod_{j\in \ri}H_{j\to i}(t),
\end{equation}
where the message $H_{j\to i}(t) = 1 - \int_0(t) \lambda e^{-(\rho+\lambda)\tau}\left[1-z\prod_{k\in \partial j \backslash i} H_{k\to j}(t-\tau)\right]d\tau$ is defined in an integration form. And the dynamical equation of $H_{j\to i}(t)$ is:
\begin{align}
    \label{eq:Newman_MP_exponential}
    \cfrac{dH_{j\to i}(t)}{dt} = \rho-(\lambda+\rho) H_{j\to i}(t) + \lambda z \prod_{k\in\partial j \backslash i} H_{k\to j}(t).
\end{align}

The function of PA and DMP will be re-iterated in the following derivation.

\section{Derivation of Susceptible-Induced Factorization}

In the derivation, we assume that the time step $\tau$ is sufficiently small that higher-order terms like $\tau^2$ are omitted. In tensors, the dimensions of indices are labeled by $\{0,1,2\}$ representing states S, I, and R respectively. 

In this section, we detail the derivation of Susceptible-Induced Factorization with factorized operator $\tT = \prod_{i\in V} \tT_{i,\ri} \Delta_{i}$, where a tensor $\tT_{i,\ri}$ encodes single-node transition $T(X_i^{t+\tau}|X_{i}^t,X_{\ri}^t)$ and copy tensors $\Delta_{i}$ as generalized Kronecker deltas that copy the contracted basis vector to all other indices. 

The contraction of $\bra{S_i}$ and $\tT_{i,\ri}$ is as follows:
\begin{equation}\label{eq:SP_SM}
\begin{aligned}
    \bra{S_i}\tT_{i,\ri} &= \tT_{\ri}^{S} \bra{S_i}\\
    &= \left(\prod_{j \in \ri} \bra{U_j}\right) \bra{S_i}
\end{aligned}
\end{equation}
The first row is straightforward since $T(S_i^{t+\tau}|X_{i}^t,X_{\ri}^t)$ is non-zero only if $X_{i}^t = S$ by the dynamic of the SIR model. And the second row $\tT_{\ri}^{S} = \prod_{j \in \ri} \bra{U_j}$ uses an approximation $1-n\lambda\tau \approx (1-\lambda \tau)^n$ that transforms elements of $\tT_{\ri}^{S}$ into production form and $\tT_{\ri}^{S}$ becomes a tensor product of $\bra{U} = \bra{S} + (1-\lambda\tau)\bra{I} +\bra{R}$

By definition, $\Delta_{i}$ contracting $\bra{S_i}$ copies this basis vector to other indices. And $\tT_{j,\partial j},j \in \ri$ gets a neighbor $i$ susceptible, the contraction reflects the probability of transition $T(X_j^{t+\tau}|X_{j}^t,X_{\partial j\backslash i}^t,S_i^t)$. $S_i^t$ has no influence on $X_j^{t+\tau}$ such that the probability and the corresponding tensors are simplified to $T(X_j^{t+\tau}|X_{j}^t,X_{\partial j\backslash i}^t)$ and $\tT_{j,\partial j \backslash i}$, respectively.

As the structure of $\tT$ corresponds to network $G$, the factorized $\tT$ after contracting $\bra{S_i}$ also corresponds to the cavity network $G \backslash i$ where node $i$ and edges connected to $i$ are removed.
Hence, the remaining part of ``susceptible factorized'' $\tT$ potentially becomes a production of independent parts, which is consistent with connected components of $G \backslash i$, since the connection through indices and edges are broken.
Hence, we collect tensors in each independent part and label them by corresponding connected components: 
\begin{equation}
    \tT_{\mathcal{C}}^{i}= \left(\prod_{k \in (\ri\cap \mathcal{C})} \bra{U_k}\right) \left(\prod_{j\in \mathcal{C}} \tT_{j,\partial j\backslash i} \Delta_{j}\right).
\end{equation}
Hence, we obtain the compact function $\bra{S_i} \tT = \prod_{\mathcal{C}\in \mathbf{C}(G\backslash i)} \tT_{\mathcal{C}}^{ i} \bra{S_i}$

Apply the factorization property on the contraction of $\bra{S_i}$ and a state $\tP(t) = \tT^{t/\tau}\tP(0)$ we have:
\begin{equation}
     \bra{S_i}\tP(t) = \left(\prod_{\mathcal{C}\in \mathbf{C}(G\backslash i)} \tT_{\mathcal{C}}^{ i}\right)^{t/\tau}\bra{S_i}\tP(0).
\end{equation}
Given an uncorrelated initial state $\tP(0) = \prod_{i \in V} \tP_{i}(0)$, in which tensor $\tP_{i}(0) = P(S_i,0)\ket{S}+ P(I_i,0)\ket{I}+ P(R_i,0)\ket{R}$ encodes an independent initial state of node $i$. Then the contraction is as follows:
\begin{equation}\label{eq:PS_SM}
    \begin{aligned}
        \bra{S_i}\tP(t) &= P(S_i,0) \prod_{\mathcal{C} \in \mathbf{C}(G\backslash i)} \tP_\mathcal{C}^i(t),\\
        \tP_\mathcal{C}^i(t) &= \left(\tT^{i}_{\mathcal{C}}\right)^{t/\tau} \prod_{j \in \mathcal{C}} \tP_{j}(0).
    \end{aligned}
\end{equation}
The tensor $\tP_\mathcal{C}^i(t)$ can be roughly viewed as the state of the component $\mathcal{C}$ evolved
under the modified operator $\tT^{i}_{\mathcal{C}}$ that is not normalized. In next two sections, we will derive DMP, PA and local update equation for our method by this function.

\section{Derivation of dynamical message passing}

As shown in the main text, the key function of DMP is obtained from calculating susceptible marginal $P(S_{i},t)$ by summarizing all other indices of $\bra{S_i}\tP(t)$, which we represent by contracting the summary vectors $\bra{\Sigma} = \bra{S} + \bra{I} + \bra{R}$:
\begin{equation}
    \label{eq:general_marginal}
    \begin{split}
        P(S_i,t) &= \left(\prod_{j\in G\backslash i}\bra{\Sigma_j}\right) \bra{S_i}\tP(t)\\
        &= P(S_i,0) \prod_{\mathcal{C} \in \mathbf{C}(G\backslash i)} \left(\prod_{j\in \mathcal{C}}\bra{\Sigma_j}\right)\tP_\mathcal{C}^i(t)\\
        &= P(S_i,0) \prod_{\mathcal{C} \in \mathbf{C}(G\backslash i)} H_{\mathcal{C}}^{i}(t),
    \end{split}
\end{equation}
where we denote the scalars by $H$ for consistency with classical DMP and to distinguish them from probabilities.
To rigorously transform production from connected components to single neighbors, connected components must consist of only one neighbor of $i$, which requires that there is no loop consisting of node $i$. Apply the condition to all nodes in the network, requiring a tree structure without loop. Such that we have the tree-specific function:
\begin{equation}
    \label{eq:MP_marginal_by_TN}
    P(S_i,t) = P(S_i,0) \prod_{j \in \ri} H_{j}^{i}(t).
\end{equation}
The label of connected components $\mathcal{C}$ replaced by consisting neighbors $j$.
We have the initial condition of scalars $H_{j}^{i}(t)$ that $H_{j}^{i}(0) = \prod_{j\in \mathcal{C}} \bra{\Sigma_j} \tP_{j}(0) = 1$ by the normalized probability. 

Calculating $H_{j}^{i}(t)$ by contracting a sub tensor network in every time step is unfeasible. Hence, dynamical equations are introduced to update $H_{j}^{i}(t)$ like existing DMP methods. The difference of $H_{j}^{i}(t)$ through a time step is:
\begin{equation}
    \label{eq:message_difference}
    \begin{aligned}
        H_{j}^{i}(t+\tau)-H_{j}^{i}(t) =& \left( \prod_{k\in \mathcal{C}_j} \bra{\Sigma_k} \tT^{i}_{\mathcal{C}_j} - \prod_{k\in \mathcal{C}_j} \bra{\Sigma_k} \right) \tP_{\mathcal{C}_j}^i(t)\\
        =& \left( \bra{U_j}\prod_{k\in \mathcal{C}_j} \bra{\Sigma_k} \tT_{k,\partial k\backslash i} \Delta_{k} - \prod_{k\in \mathcal{C}_j} \bra{\Sigma_k} \right)\tP_{\mathcal{C}_j}^i(t)\\
        =& \left( \bra{U_j}\prod_{k\in \mathcal{C}_j\backslash j} \bra{\Sigma_k} - \prod_{k\in \mathcal{C}_j} \bra{\Sigma_k} \right)\tP_{\mathcal{C}_j}^i(t)\\
        =& -\lambda \tau  \bra{I_j}\prod_{k\in \mathcal{C}_j\backslash j} \bra{\Sigma_k} \tP_{\mathcal{C}_j}^i(t)
    \end{aligned}
\end{equation}
The third row is achieved with $\prod_{k\in \mathcal{C}_j} \bra{\Sigma_k} \tT_{k,\partial k\backslash i} \Delta_{k}= \prod_{k\in \mathcal{C}_j\backslash j} \bra{\Sigma_k}$, which can be simply derived from the Markovian property of $\tT_{k,\partial k\backslash i}$ that $\sum_{X_k^{t+\tau}} P(X_k^{t+\tau}|X_{k}^t,X_{\partial k}^t)= 1$.
Hence, we have the derivative of $H_{j}^{i}(t)$ with $\tau \to 0$:
\begin{equation}
    \begin{aligned}
    \label{eq:message_dynamical_TN}
    \cfrac{dH_{j}^{i}(t)}{dt} &= \lim_{\tau \to 0}\cfrac{H_{j}^{i}(t+\tau)-H_{j}^{i}(t)}{\tau} \\
    &= -\lambda \bra{I_j}\prod_{k\in \mathcal{C}_j\backslash j} \bra{\Sigma_k} \tP_{\mathcal{C}_j}^i(t).
    \end{aligned}
\end{equation}

Then we further transform the contraction into a close equation with $H_{j}^{i}(t)$ itself. First, we use a trick that $ \bra{I} = \bra{\Sigma} - \bra{S} - \bra{R}$, then we have:
\begin{equation}
    \label{eq:message_dynamic_split}
    \cfrac{dH_{j}^{i}(t)}{dt}  = - \lambda  \left(\bra{\Sigma_j} - \bra{S_j} - \bra{R_j}\right) \prod_{k\in \mathcal{C}_j\backslash j} \bra{\Sigma_k} \tP_{\mathcal{C}_j}^i(t).
\end{equation}
Thus, there are three terms in the function: The first term of $\bra{\Sigma_j}$ is $H_{j}^{i}(t)$ itself with a factor $- \lambda$. 
Recall Eq.~\ref{eq:SP_SM}, the vector $\bra{S_j}$ can also factor $\tT_{j,\partial j\backslash i}$ and $\Delta_{j}$ in $\tT_{\mathcal{C}_j}^i$ of $\tP_j^i(t)$. The transformation of one layer of the operator $\tT_{\mathcal{C}_j}^i$ is:
\begin{equation}
\begin{aligned}
    \bra{S_j} \tT_{\mathcal{C}_j}^i &= \bra{S_j} \bra{U_j} \prod_{k\in \mathcal{C}_j} \tT_{k,\partial k\backslash i} \Delta_{k}\\
    &= \braket{U_j|S_j} \prod_{\mathcal{C}\in \mathbf{C}(\mathcal{C}_j\backslash j)} \prod_{k\in \mathcal{C}} \tT_{k,\partial k\backslash i} \Delta_{k} \bra{S_j} \\
    &= \prod_{k\in \partial j\backslash i} \tT_{\mathcal{C}_k}^j \bra{S_j}.
\end{aligned}
\end{equation}
The last row is again relied on the tree structure to divide tensors by neighbors.
The term with $\bra{S_j}$ is simplified as follows: 
\begin{equation}
    \label{eq:S_term_SM}
\begin{aligned}
    \lambda \bra{S_j} \prod_{k\in \mathcal{C}_j\backslash j} \bra{\Sigma_k} \tP_j^i(t) 
    &= \lambda P(S_j,0) \prod_{k\in \partial j\backslash i} \prod_{l\in \mathcal{C}_k} \bra{\Sigma_l}\tP_{\mathcal{C}_k}^{j}(t)\\
    &= \lambda P(S_j,0) \prod_{k\in \partial j\backslash i} P_k^j(t)
\end{aligned}
\end{equation}

For the last term, we start by contracting on one layer of operator:
\begin{equation}
\begin{aligned}
    \bra{R_j} \prod_{k\in \mathcal{C}_j\backslash j} \bra{\Sigma_k}  \tT_{\mathcal{C}_j}^i
    =& \bra{R_j}\tT_{j,\partial j\backslash i } \bra{U_j}\Delta_{j} \prod_{k\in \mathcal{C}_j\backslash j} \bra{\Sigma_k} \tT_{k,\partial k\backslash i} \Delta_{k}\\
    =&\left( \bra{R_j} +\rho \tau \bra{I_j} \right) \bra{U_j}\Delta_{j} \prod_{k\in \mathcal{C}_j\backslash j} \bra{\Sigma_k}\\
    =&\left( \bra{R_j} +\rho \tau \bra{I_j} \right)\prod_{k\in \mathcal{C}_j\backslash j} \bra{\Sigma_k}
\end{aligned}
\end{equation}
Here we employ the first order approximation of $\tau$ on $\left( \bra{R_j} +\rho \tau \bra{I_j} \right) \bra{U_j}\Delta_{j}$, which equals $\bra{R_j} +\rho \tau(1-\lambda \tau) \bra{I_j}$ and we omit the second order term.
The first term in above equation is $\bra{R_j} \prod_{k\in \mathcal{C}_j\backslash j} \bra{\Sigma_k}$ itself, and the second term with $\bra{I_j}$ corresponds to the derivative of $H_{j}^{i}(t)$. Apply this result, we have:
\begin{equation}\label{eq:R_term_SM}
\begin{aligned}
    \lambda  \bra{R_j} \prod_{k\in \mathcal{C}_j\backslash j} \bra{\Sigma_k} \tP_{\mathcal{C}_j}^i(t) 
    & = \lambda  \bra{R_j} \prod_{k\in \mathcal{C}_j\backslash j} \bra{\Sigma_k}\tP_{\mathcal{C}_j}^i(0) 
    + \lambda \sum_{T = 0}^{t-\tau}\rho \tau \bra{I_j}  \prod_{k\in \mathcal{C}_j\backslash j} \bra{\Sigma_k} \tP_{\mathcal{C}_j}^i(T)\\
    & = \lambda R_{j}(0)
    - \sum_{T = 0}^{t-\tau}\rho \tau \cfrac{dH_{j\to i}(T)}{dt}\\
    &= \lambda R_{j}(0)+ \rho \left[1 - H_{j\to i}(t)\right]
\end{aligned}
\end{equation}
Substitute Eq.~\eqref{eq:S_term_SM} and ~\eqref{eq:R_term_SM} to Eq.~\eqref{eq:message_dynamic_split},
\begin{equation}
    \label{eq:message_dynamic_final}
    \begin{aligned}
        \cfrac{dH_{j}^{i}(t)}{dt} & = -\lambda H_{j}^{i}(t) + \lambda P(S_j,0) \prod_{k\in\partial j \backslash i} H_{k}^{j}(t) + \lambda R_{j}(0) + \rho \left[1 - H_{j}^{i}(t)\right]\\
        & = \rho-(\lambda+\rho) H_{j\to i}^{T} + \lambda P(S_j,0) \prod_{k\in\partial j \backslash i} H_{k}^{j}(t) + \lambda R_{j}(0).
    \end{aligned}
\end{equation}
It is the close dynamical equation of $H_{j\to i}^{T}$, with the condition of a tree graph and uncorrelated initial states. We note that we found an extra term $\lambda R_{j}(0)$ that does not appear in the previous work, as the initial recovery probability is commonly not considered.

\section{Derivation of tensor network dynamical message passing}

Another application of Eq.~\ref{eq:PS_SM} is the joint probabilities such as $P(\mathbf{X}_A S_i,t)$, where $\mathbf{X}_A = \{X_j|j\in A\}$. It is obtained by contracting a production of projectors and sum vectors $\left(\prod_{j\in A}\bra{X_j}\right)\left(\prod_{k\in G\backslash A,i}\bra{\Sigma_k}\right)$ on $\bra{S_i}\tP(t)$:
\begin{equation}
\begin{aligned}
    P(\mathbf{X}_A S_i,t) &= P(S_i,0)\left(\prod_{j\in A}\bra{X_j}\right)\left(\prod_{k\in G\backslash A,i}\bra{\Sigma_k}\right) \prod_{\mathcal{C} \in \mathbf{C}(G\backslash i)} \tP_\mathcal{C}^i(t)\\
    &= P(S_i,0) \prod_{\mathcal{C} \in \mathbf{C}(G\backslash i)} \left(\prod_{j \in \mathcal{C} \cap A}\bra{X_j}\right) \left(\prod_{k\in \mathcal{C}\backslash A}\bra{\Sigma_k}\right) \tP_\mathcal{C}^i(t)\\
    &= P(S_i,0) \prod_{\mathcal{C} \in \mathbf{C}(G\backslash i),A \cap \mathcal{C} \ne \emptyset}H_{\mathcal{C}}^{i}(\mathbf{X}_{A},t) \prod_{\mathcal{C} \in \mathbf{C}(G\backslash i),A \cap \mathcal{C} = \emptyset} H_{\mathcal{C}}^i(t),
\end{aligned}
\end{equation}
where $H_{\mathcal{C}}^{i}(\mathbf{X}_{A},t) = \left(\prod_{j \in \mathcal{C} \cap A}\bra{X_j}\right) \left(\prod_{k\in \mathcal{C}\backslash A}\bra{\Sigma_k}\right) \tP_\mathcal{C}^i(t)$.

For the state with another area $B$, which satisfies $B \cap \mathcal{C} = \emptyset$ or $A \cap \mathcal{C} = \emptyset$ for each connected component $\mathcal{C}\in\mathbf{C}(G\backslash i)$, we have:
\begin{equation}\label{eq:XSX_SM}
\begin{aligned}
    P(\mathbf{X}_{A} S_{i} \mathbf{X}_{B},t) &= P(S_i,0) \prod_{\mathcal{C} \in \mathbf{C}(G\backslash i),A \cap \mathcal{C} \ne \emptyset}H_{\mathcal{C}}^{i}(\mathbf{X}_{A},t) \prod_{\mathcal{C} \in \mathbf{C}(G\backslash i),B \cap \mathcal{C} \ne \emptyset}H_{\mathcal{C}}^{i}(\mathbf{X}_{B},t)  \prod_{\mathcal{C} \in \mathbf{C}(G\backslash i),(A\cup B) \cap \mathcal{C} = \emptyset}H_{\mathcal{C}}^{i}(t) \\
    &= P(S_i,0) 
    \cfrac{P(\mathbf{X}_A S_i,t)}{P(S_i,0)\prod_{\mathcal{C} \in \mathbf{C}(G\backslash i),A \cap \mathcal{C} = \emptyset}H_{\mathcal{C}}^i(t)} 
    \cfrac{P(\mathbf{X}_B S_i,t)}{P(S_i,0)\prod_{\mathcal{C} \in \mathbf{C}(G\backslash i),B \cap \mathcal{C} = \emptyset}H_{\mathcal{C}}^i(t)} 
    \prod_{\mathcal{C} \in \mathbf{C}(G\backslash i),(A\cup B) \cap \mathcal{C} = \emptyset}H_{\mathcal{C}}^i(t) \\
    &= \cfrac{P(\mathbf{X}_A S_i,t)P(\mathbf{X}_B S_i,t)}{P(S_i,0)\prod_{\mathcal{C} \in \mathbf{C}(G\backslash i)}H_{\mathcal{C}}^i(t)} \\
    &= \cfrac{P(\mathbf{X}_A S_i,t)P(\mathbf{X}_B S_i,t)}{P(S_i,t)}.
\end{aligned}
\end{equation}
Then it is simple to obtain the pair approximation in SIR-like models by giving $A = \{j\},B=\{k\}$ each containing only one neighbor node of $i$. And the constrain on $A$ and $B$ is simplified to the pair of neighbors $j$ and $k$ are connected exclusively through the intermediate node $i$. And the global establishing condition is the tree structure with no loop.

For our method, we force the calculation of the state tensor $\tP_\alpha(t)$ in a local area $\alpha$, the naive calculation is 
\begin{align}\label{eq:alpha_update}
    \tP_\alpha(t+\tau) &= \tT_{\alpha,\partial \alpha} \tP_{\alpha,\partial \alpha}(t),\\
    \tT_{\alpha,\partial \alpha} &= \prod_{i\in \alpha} \tT_{i,\ri} \Delta_{i} \prod_{j\in \partial \alpha} \Delta_{j},
\end{align}
where $\tT_{\alpha,\partial \alpha}$ encoding $P\left(X_{\alpha}^{t+\tau} | X_\alpha^{t},X_{\partial \alpha}^{t}\right)$ is also factorized into tensors $\tT_{i,\ri}$ and $\Delta_{i}$. 

Start with a single node $j\in \partial \alpha$, which has only one neighbor $i \in \alpha$ and the tensor $\Delta_{j}$ becomes a unit matrix. Then the external index of $j$ in $\tT_{\alpha,\partial \alpha}$ only comes from $\tT_{i,\ri}$. We can peel off the infection from $j$:
\begin{equation}
    \label{eq:split_p_SM}
    \tT_{i,\ri} = \tT_{i, \ri\backslash j} \bra{\Sigma_j} + \lambda\tau (\ket{I_i}-\ket{S_i}) \bra{I_j}\bra{S_i} \prod_{k\in \ri \backslash j}\bra{\Sigma_k}.
\end{equation}
Apply on Eq.\eqref{eq:alpha_update}:
\begin{equation}\label{eq:Xatau}
\tP_\alpha^{t+\tau} = \left[\tT_{i, \ri\backslash j} \bra{\Sigma_j} + \lambda\tau (\ket{I_i}-\ket{S_i}) \bra{I_j}\bra{S_i} \prod_{k\in \ri \backslash j}\bra{\Sigma_k}\right] \Delta_{i} 
\prod_{k\in\alpha\backslash i} \tT_{k,\partial k} \Delta_{k} \prod_{l\in \partial \alpha\backslash j} \Delta_{l} 
\tP_{\alpha,\partial \alpha}^{t}.
\end{equation}
For simplicity, we denote $\prod_{k\in\alpha\backslash i} \tT_{k,\partial k} \Delta_{k} \prod_{l\in \partial \alpha\backslash j} \Delta_{l} $
The first term in the middle bracket is:
\begin{equation}\label{eq:noi_SM}
    \tT_{i, \ri\backslash j} \bra{\Sigma_j} \Delta_{i} \prod_{k\in\alpha\backslash i} \tT_{k,\partial k} \Delta_{k} \prod_{l\in \partial \alpha\backslash j} \Delta_{l} \tP_{\alpha,\partial \alpha}^{t} = \prod_{k\in\alpha\backslash i} \tT_{k,\partial k\backslash j} \Delta_{k} \prod_{l\in \partial \alpha\backslash j} \Delta_{l} \tP_{\alpha,\partial \alpha\backslash j}^{t},
\end{equation}
without dependence on the state of $j$. And for the second term:
\begin{equation}\label{eq:ISX_SM}
\begin{aligned}
    \bra{I_j}\bra{S_i} \Delta_{i} \tP_{\alpha,\partial \alpha}^{t} &= \prod_{m\in(\partial i\cap \alpha)}\ket{S_m}\bra{I_j}\bra{S_i} 
    \tP_{\alpha,\partial \alpha}(t)\\
    &= \prod_{m\in(\partial i\cap \alpha)}\ket{S_m}\cfrac{P(I_jS_i,t)}{P(S_i,t)} \bra{S_i}\tP_{\alpha,\partial \alpha \backslash j}(t)\\
    &= m_{j\to i}(t)\bra{S_i} \Delta_{i}\tP_{\alpha,\partial \alpha \backslash j}(t),
\end{aligned}
\end{equation}
where the other factor remains in the derivation. In this part, the message $m_{j\to i}(t) = P(I_jS_i,t)/P(S_i,t)$ efficiently replaced the dependence on the state of $j$. 
Substitute upper function and Eq.\eqref{eq:noi_SM} to Eq.\eqref{eq:Xatau}:
\begin{equation}
\begin{aligned}
    \tP_\alpha^{t+\tau} 
    &= \left[\tT_{i, \ri\backslash j} + \lambda\tau (\ket{I_i}-\ket{S_i}) m_{j\to i}(t)\bra{S_i} \prod_{k\in \ri \backslash j}\bra{\Sigma_k}\right] \Delta_{i} 
    \prod_{k\in\alpha\backslash i} \tT_{k} \Delta_{k} \prod_{l\in \partial \alpha\backslash j} \Delta_{l} 
    \tP_{\alpha,\partial \alpha \backslash j}^{t} \\
    &= \tT_{i, \ri\backslash j}(m_{j\to i}(t)) \Delta_{i} 
    \prod_{k\in\alpha\backslash i} \tT_{k} \Delta_{k} \prod_{l\in \partial \alpha\backslash j} \Delta_{l} 
    \tP_{\alpha,\partial \alpha \backslash j}^{t}.
\end{aligned}
\end{equation}
We denote $\tT_{i, \ri\backslash j}(m_{j\to i}(t))$ as the local transition tensor with index of $j$ replaced by $m_{j\to i}(t)$. 
Moreover, the influence of multiple neighbors can be simply expressed as the addition of messages as: $\tT_{i, \ri\backslash j}(m_{\partial \alpha \to i}(t))$ where $m_{\partial \alpha \to i}(t) = \sum_{j\in (\ri\cap\partial \alpha)}m_{j\to i}(t)$.

Furthermore, assuming that all external neighbors of all nodes in $\alpha$ meet the condition of the previous derivation, we can replace all the indices of $\partial \alpha$ by the corresponding messages. So we have:
\begin{equation}\label{eq:local_update_SM}
\begin{aligned}
    \tP_\alpha(t+\tau) &= 
    \prod_{i \in \alpha} \tT_{i, \ri \cap \alpha }(m_{\partial \alpha \to i}(t)) \Delta_{k} 
    \tP_{\alpha}(t) \\
    &= \tT_{\alpha}(m_{\partial\alpha}(t))  \tP_{\alpha}(t),
\end{aligned}
\end{equation}
we apply $\tT_{i, \ri \cap \alpha }(m_{\partial \alpha \to i}(t))$ on all nodes, since $m_{\partial \alpha \to i}(t) = 0$ if a node $i$ has not external neighbor.

We have obtained the local update function, but it only establishes in special areas whose external neighbors are exclusively connected to the area through a node in the area. 
There is a trivial case that the update of single node marginals are $\tP_{i}(t+\tau) = \tT_{i}(m_{\ri}(t)) \tP_{i}(t)$ which we use to calculate marginals in our method. 
In the case of an area with multiple nodes, the condition means that there is no loop crossing the eligible area, that the edges of a loop are either all contained by the area or that no edge of the loop consists of the area. 
Consequently, to find the eligible areas, we can remove all edges consisting of no loop, that each edge is a minimal eligible area, and the connected components of the left networks are other eligible areas. The areas found in such a method are the minimum units because they cannot be divided into smaller eligible areas (except single nodes) while their combination also fits the condition of Eq.~\eqref{eq:local_update_SM}. Hence we call them regions, and they are consistent with theory as biconnected components in graph.

\section{Approximate partition algorithm}

\begin{figure}[!htb]
    \centering
    \includegraphics[width=0.5\linewidth]{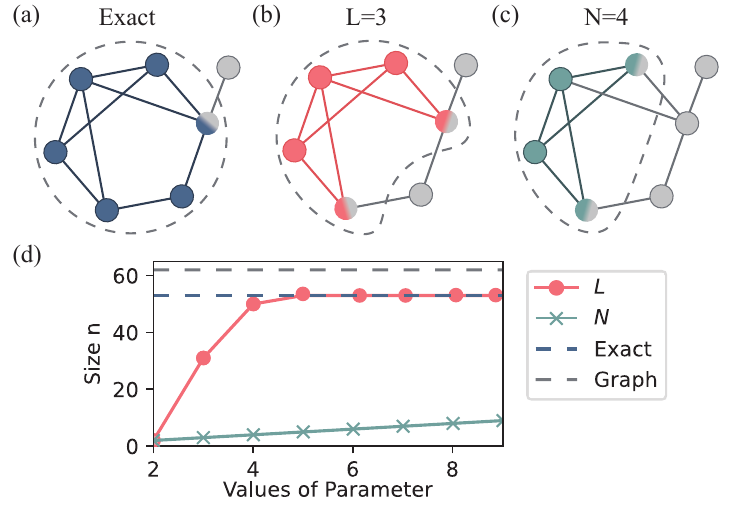}
    \caption{\textbf{Exact and approximate partition on an example network and the dolphin social network.} (a-c) An application of exact partition and approximate partition with $L$ and $N$. Dashed lines circle non-trivial regions with different colors and other gray edges are trivial regions, while the common nodes are in gradient coloring. (a): Exact partition, (b): Approximation with $L=3$, (c): Approximation with $N=4$.
    (d) The size of largest region of a real-world social network of dolphins, obtained by approximation with $L$ and $N$. The network has $n=62$ nodes and the largest exact region has a size of 53.}
    \label{fig:partition}
\end{figure}

First, we consider an approximation that ignores excessively long loops, parameterized by an upper bound for length $L$. 
In this approach, regions collect only short loops (length $\le L$), and edges that only consist of longer loops are divided into single-edge regions. 
This approximation is based on the widely accepted assumption that the error decreases with growing loop length, a principle widely used in other works to fix loop errors. 
This approximation is implemented by removing edges from exact regions on the basis of the length of the shortest loop containing them. 
An example with $L=3$ in Fig.~\ref{fig:partition}(b) reduces the region size from 6 to 5 with minimal error.

This $L$-based approximation is not an efficient constraint on region size, as it struggles to split dense-connected regions rich in short loops, as shown in Fig.~\ref{fig:partition}(d), where a dominant region remains even for $L<5$, and the smallest improvement of $L=3$ results in an intractable region size of 31. Given the impracticality of the $L$-based approximation for general networks, we introduce the approximation that directly sets a maximal region size $N$, and divides oversized regions accordingly.

In the approximation based on $N$, an oversized exact region is divided into approximate regions whose size does not exceed the parameter $N$.
This approach is also based on the negative correlation between the length of loops and the corresponding error. In analogy to the approximation with $L$, we remove edges from regions in the order of the minimum length of loops that contain them. However, in this approximation, the removal continues until the region size is sufficiently small. As a consequence, there could be complete loops in the edges removed.
By definition, the approximation with $N=\infty$ is returned to strict, as regions of any size are allowed. 
At the other extreme, $N=2$ forces each region to contain only one edge, effectively assuming no loops and degenerating TNDMP into a tensor-method equivalent to the pair approximation.

An example with $N=4$ in Fig.~\ref{fig:partition}(c) shows a stronger region truncation than the approximation with $L=3$. 
In Fig.~\ref{fig:partition}(d), the $N$-based approximation demonstrates a linear reduction in maximum region size (green line ), which has been shown to be more effective than the $L$-based approach.
Furthermore, combined approximation with two parameters $L,N$ is also feasible and have potential on more complex networks.
Note that, the homologous two approximation have a relation that approximate regions obtained with $N$ cannot contain loops longer than $N$, such that they are sub networks of the regions obtained from approximation with $L=N$. This property is also used in the algorithm.
Unlike the $L$-based method, the partition with a given $N$ is not unique; a greedy heuristic algorithm is used, with optimal partition searching left for future research.

Our algorithm of $N$-based repartitioning on an oversized regions is a recursive procedure that systematically isolates appropriately sized sub-regions. 
In one iteration, we first obtain the length of the shortest loop encompassing edge $e$, which is denoted $l_e$. Then we remove all edges with maximum $l_e$ if the remaining part has a size still larger than $N$. 
If removing present max-$l_e$ edges will make the remaining network smaller than $N$, a secondary criterion is applied. We introduce $n(l_e)$, defined as the number of distinct loops of length $l_e$ containing the edge $e$. 
Edges with a lower $n(l_e)$ are prioritized for removal, under the assumption that their exclusion introduces a smaller error by disrupting fewer long-range correlations. 
This process continues until the primary component's size is at most $N$, at which point this component is accepted as a valid $N$-approximate region.
Unlike the $L$-based method, the excised edges and their associated vertices are not discarded. Instead, they constitute a new subgraph. The non-trivial biconnected components of this new subgraph are then recursively subjected to the same $N$-based re-partitioning procedure. This recursion terminates when all remaining components are trivial (i.e., single edges or trees).

In realization, we use a trick from aforementioned property that we perform the $N$-based re-partitioning on $L$-approximate regions given $L=N$, instead of exact regions. 
Therefore, the complete procedure of a $N$-based approximation is composed of three partitioning in sequence: a biconnected component partitioning, a $L$-based repartitioning given $L=N$, and the final $N$-based repartitioning. The complete procedure is summarized in the provided pseudo-code.

\end{document}